\documentclass[]{iopart}
\usepackage[english]{babel}
\usepackage[latin1]{inputenc}
\usepackage{hyperref}

\usepackage{graphicx}
\usepackage{amsmath}
\usepackage{xcolor}
\usepackage{mathrsfs}
\usepackage{iopams} %carica in automatico i packages amsgen,amsfonts,amsbsy,amssymb
\DeclareMathOperator{\Id}{ {\bf 1}}
\newcommand{\map}[1]{\mathscr{#1}}
\newcommand{\cnot}{\text{CNOT}}
\newcommand{\cphase}{\text{CZ}}

\usepackage[matrix,frame,arrow]{xy}
\usepackage{amsmath}
\newcommand{\bra}[1]{\left\langle{#1}\right\vert}
\newcommand{\ket}[1]{\left\vert{#1}\right\rangle}
\newcommand{\qw}[1][-1]{\ar @{-} [0,#1]}
\newcommand{\qwx}[1][-1]{\ar @{-} [#1,0]}
\newcommand{\gate}[1]{*{\xy *+<.6em>{#1};p\save+LU;+RU **\dir{-}\restore\save+RU;+RD **\dir{-}\restore\save+RD;+LD **\dir{-}\restore\POS+LD;+LU **\dir{-}\endxy} \qw}
\newcommand{\control}{*!<0em,.025em>-=-{\bullet}}
\newcommand{\ctrl}[1]{\control \qwx[#1] \qw}
\newcommand{\multigate}[2]{*+<1em,.9em>{\hphantom{#2}} \qw \POS[0,0].[#1,0];p !C *{#2},p \save+LU;+RU **\dir{-}\restore\save+RU;+RD **\dir{-}\restore\save+RD;+LD **\dir{-}\restore\save+LD;+LU **\dir{-}\restore}
\newcommand{\ghost}[1]{*+<1em,.9em>{\hphantom{#1}} \qw}
\newcommand{\Qcircuit}[1][0em]{\xymatrix @*=<#1>}
\begin{document}

\title{Noise robustness in the detection of non separable random unitary maps}

\author{C. Macchiavello and M. Rossi}
\address{Dipartimento di Fisica and INFN-Sezione di Pavia, 
via Bassi 6, 27100 Pavia, Italy}

\date{\today}
\begin{abstract}
We briefly review a recently proposed method to detect properties of quantum 
noise processes and  quantum channels.
We illustrate in detail the method for detecting
non separable random unitary channels and consider in particular the explicit
examples of the $\cnot$ and $\cphase$ gates. We analyse their robustness in the 
presence of noise for several quantum noise models.    
\end{abstract}

%\maketitle

\section{Introduction}

Quantum noisy channels, and in general quantum noise processes, can be 
measured by means of complete process tomography \cite{NC}. Tomography does not need
any a priori knowledge about the quantum process under consideration but
at the same time it requires a large number of measurement settings when
it has to be implemented experimentally (which goes as $d^4$, where $d$ is 
the dimension of the quantum system on which the channel acts).
In many realistic implementations, however, some a priori information  
on the form of a quantum channel, 
or a quantum noise process, is available and it is of great interest to 
determine experimentally with the minimum number of measurement
settings whether or not the channel
has a certain property (e.g. being entanglement breaking or non separable random unitary). In this work we review a recently proposed efficient 
method for quantum channel detection \cite{mapdet} by avoiding complete 
quantum process tomography
and apply it to non separable random unitary channels.
In particular, we study in detail its robustness in the presence of noise.

The present paper is organised as follows. In Sect. \ref{s:preli} we remind
some preliminary notions that represent the main ingredients to develop
the proposed 
quantum channel detection method, namely the  Choi-Jamolkowski isomorphism
and the entanglement witnesses. In Sect. \ref{s:RU} we illustrate
the method in the case of detection of non separable random unitary maps. 
In Sect. \ref{s:noise} we study in detail the robustness of the method in the 
presence of noise for depolarising, dephasing, bit flip and 
amplitude damping noise.
In Sect. \ref{s:conc} we finally summarise the main results.

\section{Preliminaries}
\label{s:preli}

Quantum channels, and in general quantum noise processes, 
are described by completely positive and trace preserving 
(CPT) maps $\map{M}$, which can be expressed in the Kraus form \cite{kraus}
as
\begin{equation}
\map{M}[\rho]=\sum_k A_k\rho A_k^\dagger,
\end{equation}
where $\rho$ is the density operator of the quantum system on which the 
channel acts and the Kraus operators $\{A_k\}$ fulfil the constraint 
$\sum_k A_k^\dagger A_k=\Id$. 

In order to develop the detection method proposed, we will use the 
Choi-Jamolkowski isomorphism \cite{jam, choi}, which gives a one-to-one
correspondence between CPT maps acting on $\mathcal{D(H)}$ (the set of 
density operators on $\mathcal{H}$) and bipartite density operators 
$C_{\map{M}}$ on $\mathcal{H\otimes H}$. This isomorphism can be described as
\begin{equation}
\map{M}\Longleftrightarrow 
C_{\map{M}}=\map{M}\otimes\map{I}[\ket{\alpha}\bra{\alpha}],
\end{equation}
where $\map{I}$ is the identity map, 
and $\ket{\alpha}$ is the maximally entangled state with respect to the 
bipartite space $\mathcal{H\otimes H}$, i.e. $\ket{\alpha}=\frac{1}{\sqrt 
d}\sum_{k=1}^d\ket{k}\ket{k}$ (we consider here quantum channels acting 
on systems with finite dimension $d$).

By exploiting the above isomorphism, we are able to link some specific 
properties 
of quantum channels to properties of the corresponding Choi states 
$C_{\map{M}}$. In particular, we find a connection between quantum channel 
properties and (multipartite) entanglement properties of the corresponding 
Choi states. The method works when we consider properties that are based on a 
convex structure of the quantum channels. 

\begin{figure}[h!]
\centering
\includegraphics[scale=.5]{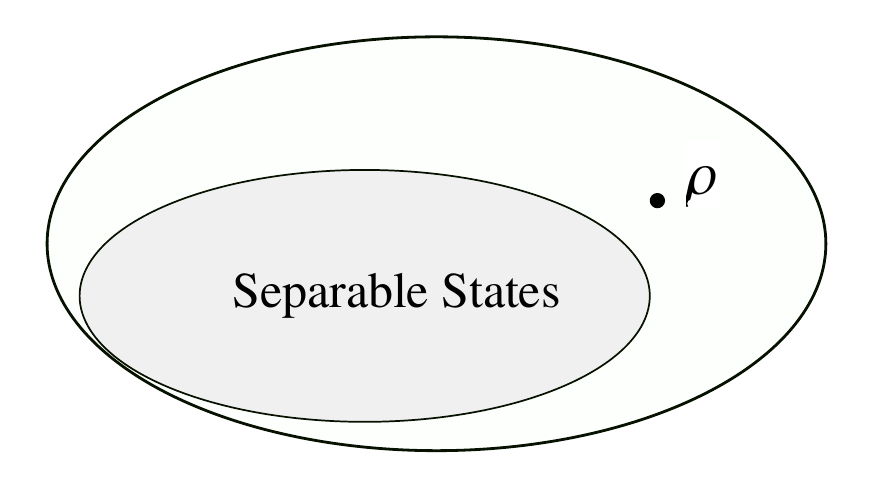}
\includegraphics[scale=.5]{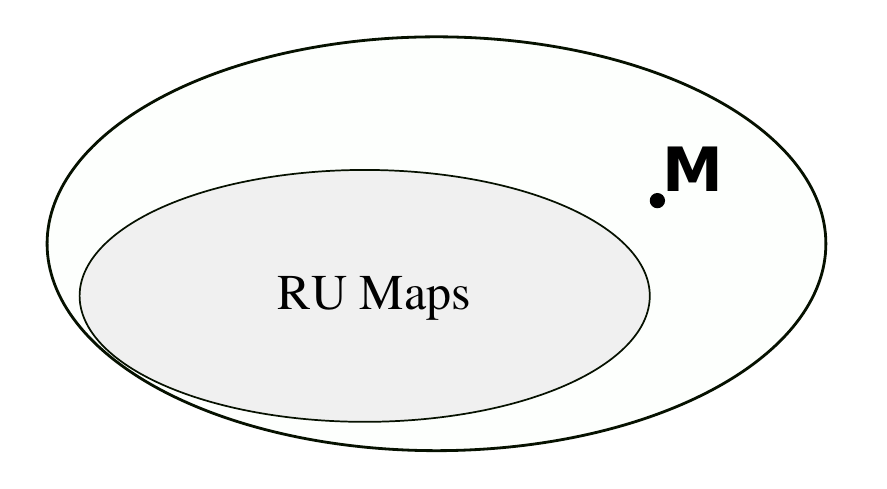}
\caption{Comparison between the sets of quantum states and quantum maps. 
The set of random unitary channels is denoted by RU, and it is defined in 
Eq. \eqref{RU} below.}
\label{sets}
\end{figure}

The second main ingredient that is employed is the concept of
entanglement detection via witness operators \cite{horo-ter}. 
We then briefly remind here that a state $\rho$ is entangled 
if and only if there exists a hermitian operator $W$ such that 
$\Tr[W\rho]< 0$ and $\Tr[W\rho_{sep}]\geq 0$ for all separable states.
The correspondence that we exploit is between the Choi states of the considered
set of quantum channels and the set of separable states. 
Both represent convex subsets of the sets
of all quantum channels acting on density operators on $\mathcal{H}$
and all bipartite density operators on $\mathcal{H\otimes H}$ respectively, 
as shown in Fig. \ref{sets}.

\section{Non separable random unitary maps}
\label{s:RU}

We will now illustrate explicitly how the channel detection method works
in the case of separable random unitary maps. Let us first remind the concept 
of random unitary channels (RU). These are defined as 
\begin{equation}\label{RU}
\map{U}[\rho]=\sum_k p_k U_k\rho U_k^\dagger,
\end{equation}
where $U_k$ are unitary operators and $p_k\ge 0$ with $\sum_k p_k=1$. Notice 
that this kind of maps includes several interesting models of quantum noisy
channels, such as the depolarising channel or 
the phase damping channel and the bit flip channel \cite{NC}. 

Let us now assume that the system on which the random unitary channel acts
is a bipartite system $\rho_{AB}$ (composed of systems A and B).
We can then identify a class of random unitary maps which is separable, 
namely that can be written in the form
\begin{equation}\label{SRU}
\map{V}[\rho_{AB}]=\sum_k p_k (V_{k,A}\otimes W_{k,B})\rho_{AB} (V_{k,A}^\dagger
\otimes W_{k,B}^\dagger),
\end{equation}
where $\rho_{AB}$ is a bipartite system, and both $V_{k,A}$ and $W_{k,B}$  
are unitary operators for all $k$'s, acting on systems A and B respectively. 
Quantum channels of the
above form are named separable random unitaries (SRU) and they form a convex
subset in the set of all CPT maps acting on bipartite systems $\rho_{AB}$. 
Interesting examples of channels of this form are given by Pauli memory 
channels \cite{memory}.

When considering quantum channels acting on bipartite systems, 
the Choi state is a four-partite state (composed of systems A, B, C and D).
Notice that the state 
$\ket{\alpha}=\frac{1}{\sqrt{d_{AB}}}\sum_{k,j=1}^{d_{AB}}
\ket{k,j}_{AB}\ket{k,j}_{CD}$ (where $d_{AB}=d_Ad_B$ 
is now the dimension of the 
Hilbert space of the bipartite system AB) can also be written as  
$\ket{\alpha}=\ket{\alpha}_{AC}
\ket{\alpha}_{BD}$, namely it is a biseparable state for the partition AC|BD
of the global four-partite system.
The Choi states corresponding to SRU channels therefore form a convex set, 
which is a subset of all biseparable states for the partition AC|BD.
Since the generating set of separable random unitaries is
given by local unitaries $U_A\otimes U_B$,  
the generating bipartite pure states in the corresponding convex
set of Choi states have the form (we name this set of four-partite density 
operators $S_{SRU}$)
\begin{equation}\label{ext}
(U_A\otimes\Id_{C})\ket{\alpha}_{AC}\otimes(U_B\otimes\Id_{D})
\ket{\alpha}_{BD}\;.
\end{equation}
We can now detect non separable RU maps (which correspond to Choi states
that are entangled in the bipartition AC|BD) by designing suitable
witness operators that detect the corresponding Choi state with
respect to biseparable in AC|BD states belonging to $S_{SRU}$. 

We illustrate this procedure with a simple example. Consider the case of 
detecting a non separable unitary operation $U$ acting on a bipartite system 
AB. A suitable detection operator can be constructed as 
\begin{equation}\label{W}
W_U=\beta\Id-C_U\;,
\end{equation}
where the coefficient $\beta$ is the squared overlap between the closest 
biseparable state in the set $S_{SRU}$ and the entangled state $C_U$, 
namely
\begin{equation}\label{beta}
\beta =\max_{\ket{\phi}\in S_{SRU}}\bra{\phi}C_U\ket{\phi}.
\end{equation}
Notice that, since the maximum of a linear function over a convex set is 
always achieved on the extremal points, the maximum above can be always 
calculated by maximising over the pure biseparable states (\ref{ext}) 
\cite{gen}.

We will specify the above construction to the particular case of the
CNOT gate acting on two qubits. The corresponding Choi state has the form
\begin{equation}\label{cnot}
C_\cnot =(\cnot\otimes\Id)\ket{\alpha}\bra{\alpha}(\cnot\otimes\Id),
\end{equation}
where the CNOT operation is given by 
\begin{equation}
\cnot=
\begin{pmatrix}
\Id & 0 \\
0 & X 
\end{pmatrix}\;,
\end{equation}
with $\Id$ representing the $2\times 2$ identity matrix, and $X$ the 
Pauli operator $\sigma_x$. 

The optimal coefficient $\beta$ equals $1/2$ and the detection operator 
$W_\cnot$ can be decomposed 
into a linear combination of local operators as follows \cite{mapdet}
\begin{align}\label{WCNOT}
W_\cnot =\frac{1}{64}(&31\Id\Id\Id\Id -\Id X\Id X-XXX\Id -X\Id XX \nonumber \\
				&-ZZ\Id Z+ZY\Id Y+YYXZ+YZXY\nonumber \\
				&-Z\Id Z\Id -ZXZX+YXY\Id +Y\Id YX\nonumber \\
				&-\Id ZZZ+\Id YZY + XYYZ+XZYY)\;,
\end{align}
where for simplicity of notation $X$, $Y$ and $Z$ represent 
the Pauli operators and the tensor product symbol has been omitted.
As we can see from the above form, the CNOT can be detected by using 
nine different local measurements settings \cite{jmo}.
Following \cite{ent-wit, gu-hy},  it can be also easily proved that the 
above form is optimal in the sense that 
it involves the smallest number of  measurement settings.
From the point of view of implementations, 
the optimal detection procedure then works
as follows: prepare a four-partite qubit system in the state  
$\ket{\alpha}=\ket{\alpha}_{AC}\ket{\alpha}_{BD}$, input qubits A and B to
the quantum channel and finally perform the set of nine local measurements
reported above on the four-partite system $ABCD$ in order to measure the 
operator (\ref{WCNOT}). 
If the resulting average value is negative then the quantum channel is 
detected as a non separable random unitary map.

As a second significant example consider the $\cphase$ operation, which also 
represents an important two-qubit gate in quantum computation \cite{NC}.
This operation is defined as 
\begin{equation}
\cphase=
\begin{pmatrix}
\Id & 0 \\
0 & Z 
\end{pmatrix}\;,
\end{equation}
namely it has the same structure as the CNOT gate, with $X$ replaced by $Z$.
This case can be connected to the detection procedure for the CNOT gate
by exploiting the following relation between the CNOT and $\cphase$ gates 
\begin{equation}
\cphase=(\Id\otimes H)\cnot (\Id\otimes H),
\end{equation}
where $H$ is the Hadamard gate, defined as $H=\frac{1}{\sqrt 2}(X+Z)$. Since the two gate operations differ only by a local unitary transformation,
the maximisation performed in Eq. (\ref{beta}) leads to the same value for 
$\beta$. The corresponding detection operator $W_\cphase$ 
can then be written in the form
\begin{align}\label{WCPHASE}
W_\cphase =\frac{1}{64}(&31\Id\Id\Id\Id -\Id Z\Id Z-Z\Id Z\Id - ZZZZ \nonumber \\
				- & ZX\Id X + ZY\Id Y-\Id XZX +\Id YZY \nonumber \\
				- & XZX\Id -X\Id X Z +YZY\Id + Y\Id YZ \nonumber \\
				- & YYXX -YXXY -XYYX -XXYY)\;,
\end{align}
which again corresponds to a set of nine local measurements.

\section{Noise robustness}
\label{s:noise}

We will now study the robustness of the method in the presence of additional
noise, which can influence the operation of the quantum channel.
The situation we have in mind is the following. Suppose we are given a 
witness $W_U$ of the form (\ref{W}) 
to detect a unitary transformation $U$ 
acting on two qubits. Suppose also that the 
experimental implementation of $U$ leads to a new map $\map{M}$, which is 
close to the original $U$ by construction but not exactly $U$ due to the 
presence of noise. Does the witness $W_U$ still detect the map $\map{M}$ as a 
non SRU map? To answer this question we have to check whether 
the expectation value 
of the witness $W_U$ on the map $\map{M}$ is still negative. 

Starting from the definition (\ref{W}), the expectation value of $W_U$ 
on $C_\map{M}$ can be expressed as
\begin{equation}
\Tr[W_UC_\map{M}]=\beta-\Tr[C_\map{M}C_U].
\end{equation}
By exploiting the Choi-Jamolkowski isomorphism, the overlap between two  
states $C_\map{L}$ and $C_\map{M}$ corresponding to the maps $\map{L}$ and 
$\map{M}$ acting on $\mathcal{D(H)}$ can be generally written as
\begin{equation}
\Tr[C_\map{M}C_\map{L}]=\frac{1}{d^2}
\sum_{i,j=1}^{d}\Tr[\map{M}(\ket{i}\bra{j})\map{L}(\ket{j}\bra{i})]\;,
\end{equation}
where $\{\ket{i}\}$ represents the computational basis for the Hilbert space acting on $\mathcal{H}$ with arbitrary finite dimension $d$.
In terms of the Kraus operators $\{A_k\}$ and $\{B_l\}$ of the maps $\map{M}$ and $\map{L}$ respectively, the above expression can be written as
\begin{equation}
\Tr[C_\map{M}C_\map{L}]=\frac{1}{d^2}\sum_{k,l}|\Tr[A_k^\dagger B_l]|^2,
\end{equation}
where the double summation is over the Kraus operators and 
the absolute value comes from the identity $\Tr[A^\dagger]=\Tr[A]^*$. 

In the present case, $\mathcal{H}$ is  a two qubit system of dimension $d=4$ and $\map{L}$ given by a unitary operation $U$. Therefore, the above expression takes the form
\begin{equation}\label{trmap}
\Tr[C_\map{M}C_U]=\frac{1}{16}\sum_{k}|\Tr[A_k^\dagger U]|^2,
\end{equation}
where the summation is now performed just over the Kraus operators $\{A_k\}$ of 
$\map{M}$.
The expectation value for the witness $W_U$ detecting the gate $U$ can then
be rewritten as
\begin{equation}\label{expect}
\Tr[W_U C_\map{M}]=\beta-\frac{1}{16}\sum_k|\Tr[A_k U^\dagger]|^2.
\end{equation}
In this case, the general map $\map{M}$ will thus represent a noisy 
implementation of the unitary $U$ by considering no longer a noiseless gate but 
adding some quantum noise such as the depolarising, the 
dephasing, the bit flip or the amplitude damping noise. In the following subsections 
we will treat these four different channels, and derive some bounds on the 
amount of noise that the witnesses $W_\cnot$ and $W_\cphase$, constructed to detect gates $\cnot$ and $\cphase$ respectively, can tolerate.

\subsection{Depolarising noise}
\label{s:depol}

We will consider first the case of depolarising noise $\map{D}$, 
whose action 
is described by a random unitary map of the following form 
\begin{equation}\label{dep-channel}
\Gamma_{\{p\}}[\rho]= \sum^3_{i=0}{p_i \sigma_i \rho \sigma_i}
\end{equation}
where $\sigma_0=\Id$ is the identity operator, and $\{\sigma_i\}$ ($i=1,2,3$) 
are the three Pauli operators  
$\sigma_1=X ,\sigma_2=Y, \sigma_3=Z$ respectively.
In the case of depolarising noise we have 
$p_0=1-3q/4$ (with $p\in[0,1]$), while $p_i=q/4$ for $i=1,2,3$, and therefore the parameter $q$ uniquely describes the depolarising channel.

The presence of noise in the general scenario of a controlled
$\text{C-}U_t$ unitary operation can be depicted as follows
\begin{equation}\label{depolarising}
\begin{matrix}
\Qcircuit
@C=1em @R=1.em 
{
 & \multigate{1}{\map{M}_{D,U_t}}  &  \qw \\
 & \ghost{\map{M}_{D,U_t}}  &  \qw
}
\end{matrix}
=
\begin{matrix}
\Qcircuit
@C=1em @R=.5em 
{
 & \gate{\map{D}_1} & \ctrl{1}  &  \gate{\map{D}_2}&  \qw \\
 & \gate{\map{D}_1} & \gate{U_t}  &   \gate{\map{D}_2}& \qw
}
\end{matrix}
\end{equation}
where $U_t$ is the unitary operation acting on the target qubit 
(in the cases of the CNOT and C-Z gates it is given by $X$ and 
$Z$ respectively), and each channel $\map{D}_i$ involves the 
parameter $q_i$. Notice that $q_1$ and $q_2$ are related to the depolarising 
channels $\map{D}_1$ and $\map{D}_2$, respectively. Obviously, the Kraus 
operators of the tensor product map 
$\map{D}_i\otimes\map{D}_i$, i.e. $\{D_k^i\}$, are given by the tensor 
product of the corresponding Kraus operators of the single qubit depolarising 
channel. Notice that the global resulting channel 
shown above is still a random unitary channel.

We will first start from the detection of noisy CNOT gate via the witness 
operator $W_\cnot$. 
From Eq. (\ref{trmap}), we can compute the overlap between the 
noiseless Choi state 
$C_\cnot$ and the noisy case ${\map{M}_{D,X}}$, 
where $\map{M}_{D,X}$ is the composite 
map given by (\ref{depolarising}) with $U_t=X$, as
\begin{equation}
\Tr[C_{\map{M}_{D,X}}C_\cnot]
=\frac{1}{16}\sum_{k,l}|\Tr[D^1_k\cnot D^2_l\cnot]|^2,
\end{equation}
where $\{D^1_k\}$ and $\{D^2_l\}$ are the Kraus sets of 
$\map{D}_1\otimes\map{D}_1$ and $\map{D}_2\otimes\map{D}_2$, respectively. 
By performing the calculation explicitly and remembering that, 
apart from the parameters $q_i$, 
the term on the right hand side above is a symmetric matrix in $k,l$, we 
arrive at the following expression for the expectation value
\begin{align}
\Tr[W_\cnot & C_{\map{M}_{D,X}}]=\frac{1}{2}\label{thres_dep}\\
&-\frac{1}{16}(16\bar q_1^2\bar q_2^2
+2q_1\bar q_1q_2\bar q_2
+q_1^2q_2\bar q_2
+q_1\bar q_1 q_2^2 
+\frac{5}{16}q_1^2q_2^2),\nonumber
\end{align}
with the definition $\bar q_i=1-\frac{3q_i}{4}$ for $i=1,2$. 

Let us now study some special cases of the above situation.
Suppose first that $q_2=0$, so that the noise affects the channel only 
before the $\cnot$. In this case the expectation value becomes
\begin{equation}
\Tr[W_\cnot C_{\map{M}_{D,X}}]=\frac{1}{2}-\bar q_1^2,
\end{equation}
which is negative for $q_1<\frac{4-2\sqrt2}{3}\simeq 0.39$. 
Therefore, the values of $q_1$ below this threshold lead to a detection of 
the $\cnot$ gate as a non separable random unitary. Since the situation is 
symmetric, the same obviously holds when $q_1=0$ and we are looking at $q_2$, 
namely the action of the depolarising channel either before or after the
CNOT operation leads to the same result. 
Another interesting situation is when both the channels before and after the 
$\cnot$ gate introduce the same level of noise, 
namely when $q_1=q_2=q$. In this case we get the 
following expression for the expectation value
\begin{equation}
\Tr[W_\cnot C_{\map{M}_{D,X}}]=\frac{1}{2}-\frac{1}{16}(q-2)^2(5q^2-8q+4).
\end{equation}
The $\cnot$ gate is thus detected as a non-separable random unitary map when 
$q<0.21$. Notice that the threshold in this case is not as high as the one we 
obtained before, since the situation is much noisier because there are two 
sources of noise.

We will now consider the case of the C-Z gate. 
The detection of noisy $\cphase$ gate via the witness $W_\cphase$ turns out 
to give the same threshold of noise as for the $\cnot$ gate. 
This is basically due to the symmetry properties of the depolarising noise, 
which acts isotropically along the three directions of the Pauli matrices. 
It is then straightforward to find that the expectation value of $W_\cphase$ 
on $C_{\map{M}_{D,Z}}$, namely 
$\Tr[W_\cphase C_{\map{M}_{D,Z}}]$ is exactly given by 
Eq. \eqref{thres_dep}. Hence, the analysis we performed in that case still holds 
for the $\cphase$ gate.

As we can see, the presence of local depolarising noise thus affects the 
CNOT and $\cphase$ operations in such a way that, beyond a certain amount of 
noise, the noisy CNOT and $\cphase$ operations become separable, and are no 
longer detected by our method. 

\subsection{Dephasing noise}
\label{s:deph}

Let us now assume that phase damping noise is present, acting independently 
on the two qubits A and B in general both before
and after the operation we want to detect (either CNOT or $\cphase$), 
as for the case of the depolarising noise considered above. 
Phase damping noise is described by a CPT map of the form
(\ref{dep-channel}) where the probabilities are given by
$p_0=1-q$, $p_1=p_2=0$ and $p_3=q$. Notice that also in this case
the global resulting channel is still a random unitary channel. 

In order to quantify the noise robustness of the witness $W_\cnot$ with 
respect to phase damping noise, we calculate the expectation value 
of $W_\cnot$ given by (\ref{W}) (with $\beta=1/2$) 
with respect to the 
state $C_{\map{M}_{P,X}}$, i.e. the Choi state 
corresponding to the composite map 
$\map{M}_{P,X}=(\map{P}_2\otimes\map{P}_2)\cnot(\map{P}_1\otimes\map{P}_1)$.
The problem thus reduces to evaluate the overlap between the Choi states 
$C_\cnot$ and $C_{\map{M}_{P,X}}$. By using Eq. (\ref{expect}),
this procedure leads to
\begin{equation}\label{thres_dephase}
\Tr[W_\cnot C_{\map{M}_{P,X}}]
=\frac{1}{2}-[(1-q_1)^2(1-q_2)^2+q_1q_2(1-q_1q_2)].
\end{equation}

From the above expression we can see that $\Tr[W_\cnot C_{\map{M}_{P,X}}]<0$ 
for 
certain intervals of the noise parameters $q_1$ and $q_2$. From 
the symmetry of the above expression, the action of dephasing noise either
before or after the CNOT gate leads to the same result. In this case,
namely $q_2=0$, the expectation value of $W_\cnot$ is negative for 
$q_1<1-\frac{1}{\sqrt 2}\simeq0.29$. When the dephasing 
channels introduce the same level of noise ($q_1=q_2=q$) the expectation value
of $W_\cnot$ turns our to be negative for $q<0.17$ and therefore the CNOT 
operation can be detected in this range. 

Regarding the robustness of the witness operator $W_\cphase$, we need to 
compute the expectation value of $W_\cphase$ with respect to the Choi state 
$C_{\map{M}_{P,Z}}$, representing the noisy implementation of the $\cphase$ 
gate, i.e. $\map{M}_{P,Z}=(\map{P}_2\otimes\map{P}_2)
\cphase(\map{P}_1\otimes\map{P}_1)$. Following the same calculation as before 
we get
\begin{equation}
\Tr[W_\cphase C_{\map{M}_{P,Z}}]
=\frac{1}{2}-(1-q_1-q_2+2q_1q_2)^2,
\end{equation}
which differs from the expectation value calculated for the $\cnot$ gate, 
see Eq. \eqref{thres_dephase}.

Also in this case, if the noise is present just before or after the gate, 
namely $q_2=0$ or $q_1=0$ respectively, our method detects the noisy 
$\cphase$ as a non separable random unitary map if $q_1<1
-\frac{1}{\sqrt{2}}\simeq 0,29$ (or $q_2<1-\frac{1}{\sqrt{2}}\simeq 0,29$).
This threshold is exactly the same as the one found for $W_\cnot$, thus the 
witness for detecting the $\cphase$ turns out to be as robust against 
dephasing noise as $W_\cnot$ revealing $\cnot$. If the two sources of noise 
have the same strength, i.e. $q_1=q_2=q$, then the expectation value turns 
out to be negative if the noise level is $q<\frac{1}{2}(1-\sqrt{1-\sqrt 2})
\simeq 0.18$ or $q>\frac{1}{2}(1+\sqrt{1-\sqrt 2})\simeq 0.82$. This behaviour
may seem to be  very surprising, since it follows that the witness $W_\cphase$ 
can tolerate not only low levels of noise but high levels too. The only regime where it fails is when the noise has a medium strength. This effect can be explained by noticing that dephasing noise always 
commutes with the $\cphase$ gate, thus the noise can be thought to be applied 
twice before the regarded gate. For high noise level $q$, the action of two 
consecutive dephasing processes leads almost to the identical map, since 
$Z^2=\Id$, and so the scenario can be thought as noiseless. We want to stress that this result is completely different 
from the one obtained for $W_\cnot$ since there only a low amount of noise 
was tolerated.

\subsection{Bit flip noise}

Another interesting model of noise is given by bit flip noise $\map{B}$, 
defined as a CPT map of the form \eqref{dep-channel} with probabilities 
$p_0=1-q$, $p_1=q$ and $p_2=p_3=0$. As before, 
we consider the situation in which the noise acts independently on the two 
qubits both before and after the controlled operation 
(either $\cnot$ or $\cphase$) we aim to detect.

Let us first focus on the detection of the $\cnot$ gate by the operator $W_\cnot$. 
By exploiting Eq. \eqref{expect}, where the composite map is now given by 
$\map{M}_{B,X}=(\map{B}_2\otimes\map{B}_2)\cnot(\map{B}_1\otimes\map{B}_1)$, 
we arrive at the following expectation value of $W_\cnot$ over its noisy 
implementation $\map{M}_{B,X}$
\begin{equation}
\Tr[W_\cnot C_{\map{M}_{B,X}}]
=\frac{1}{2}-[(1-q_1)^2(1-q_2)^2+q_1q_2(1-q_1q_2)].
\end{equation}
This turns out to be the same expectation value as for the case 
of dephasing noise, therefore the discussion already done below Eq.  
\eqref{thres_dephase} still holds.

In order to  study the robustness of $W_\cphase$ to detect $\cphase$ with 
additional bit flip noise, we have to evaluate the quantity 
$\Tr[W_\cphase C_{\map{M}_{B,Z}}]$ with $\map{M}_{B,Z}
=(\map{B}_2\otimes\map{B}_2)\cphase(\map{B}_1\otimes\map{B}_1)$. By using Eq. \eqref{expect}, we get
\begin{equation}
\Tr[W_\cphase C_{\map{M}_{B,Z}}]=\frac{1}{2}-(1-q_1)^2(1-q_2)^2,
\end{equation}
which allows us to derive different thresholds for the noise tolerance of 
$\cphase$. If noise is neglected either after ($q_2=0$) or before ($q_1=0$) 
the $\cphase$ gate, then the method is able to tolerate a level of noise up to 
$1-\frac{1}{\sqrt 2}$, i.e. either $q_1<1-\frac{1}{\sqrt 2}$ or 
$q_2<1-\frac{1}{\sqrt 2}$. In the case where both the noise sources show the 
same amount of noise, namely  $q_1=q_2=q$, it follows that the $\cphase$ gate 
is detected as long as $q<0.16$.

\subsection{Amplitude damping noise}
\label{s:adc}

As a last noise model we consider the amplitude damping channel,
which is not a random unitary noise and it is described by 
the following Kraus operators acting on a qubit state
\begin{equation}
A_1=\begin{pmatrix}
1 & 0 \\
0 & \sqrt{1-\gamma}
\end{pmatrix},
A_2=\begin{pmatrix}
0 & \sqrt \gamma \\
0 & 0
\end{pmatrix},
\end{equation}
where $\gamma$ is the parameter characterising the amount of damping. 

In the case of $W_\cnot$, following the same procedure described above 
and by considering now the composite map $\map{M}_{A,X}=
(\map{A}_2\otimes\map{A}_2)\cnot(\map{A}_1\otimes\map{A}_1)$, we have
\begin{align}
\Tr[W_\cnot & C_{\map{M}_{A,X}}]=
\frac{1}{2}\nonumber\\
&-\frac{1}{16}\big[(1+\sqrt{\bar\gamma_1\bar\gamma_2}(1+\sqrt{\bar\gamma_1}+\sqrt{\bar\gamma_2}))^2+\gamma_1\bar\gamma_1\gamma_2\bar\gamma_2\big],
\end{align}
where we have defined $\bar\gamma=1-\gamma$.
As in the previous cases the above expression is symmetric under exchange 
of $\gamma_1$ and $\gamma_2$. When noise acts only either before or after
the CNOT gate, e.g. $\gamma_2=0$, the above expression is negative for
$\gamma_1<0.53$.
For the particular case of $\gamma_1=\gamma_2=\gamma$ we have that the above 
expression reduces to
\begin{equation}
\Tr[W_\cnot C_{\map{M}_{A,X}}]=
\frac{1}{2}-\frac{1}{16}\big[(1+\bar\gamma(1+2\sqrt{\bar \gamma}))^2+\gamma^2\bar\gamma^2 \big],
\end{equation}
which is negative for $\gamma<0.31$. Therefore the composite map
can be detected as a non separable random unitary in this range of noise 
parameter $\gamma$.

The noise robustness of $W_\cphase$  with respect to the amplitude damping 
noise can be studied starting from the expectation value of $W_\cphase$ over 
$\map{M}_{A,Z}=(\map{A}_2\otimes\map{A}_2)\cphase(\map{A}_1\otimes\map{A}_1)$, which is given by
\begin{equation}
\Tr[W_\cphase C_{\map{M}_{A,Z}}]=\frac{1}{2}-\frac{1}{16}
(1+\sqrt{\bar \gamma_1 \bar \gamma_2 })^4.
\end{equation}
As we can see from the above expression, when noise is present only before 
the $\cphase$ gate, i.e. $\gamma_2=0$, a negative result is found for 
$\gamma_1<0.53$, exactly as for $W_\cnot$. Notice that, since the above 
expectation value is still invariant under exchange of $\gamma_1$ and 
$\gamma_2$, the same holds if noise acts just after the controlled gate. 
When noise before and after the $\cphase$ gate is the same, i.e. 
$\gamma_1=\gamma_2=\gamma$, it is easy to show that 
\begin{equation}\label{root}
\Tr[W_\cphase C_{\map{M}_{A,Z}}]=\frac{1}{2}-\frac{1}{16}(1+\bar\gamma)^4.
\end{equation}
Thus the witness operator $W_\cphase$ detects the noisy $\cphase$ as a non 
random unitary map only if $\gamma<0.31$. We would like to stress that this 
value is the same as before only because we truncate the root of Eq. 
\eqref{root} at the second digit.

\section{Conclusions}
\label{s:conc}

In summary, we have reviewed an experimentally feasible method to detect
specific properties of noisy quantum channels and we have analysed in
particular the case of detection of non separable random unitary maps.
The advantage of the present method over standard quantum process tomography
is that a much smaller number of measurement settings is needed in an 
experimental implementation. 
Moreover, the proposed scheme relies on the
implementation of local measurements and it is achievable with current 
technology, for example in a quantum optical set-up \cite{exp}.
We have also studied in detail the robustness of
the method in the presence of noise and imperfections in the channel operation
for the case of a unitary channel, considering the explicit examples of 
CNOT and $\cphase$ gates. We have discussed in particular four realistic 
noise models, namely the depolarising, the dephasing, the bit flip and the 
amplitude damping noise, and derived the corresponding noise intervals in which the method works.


\begin{thebibliography}{10}
\bibitem{gen} 
The states given by \eqref{ext} correspond to the generating points of the 
set of SRU channels, which contains the extremal points.
\end{thebibliography}
\end{document}